\documentclass[conference]{IEEEtran}
\IEEEoverridecommandlockouts
\usepackage{cite}
\usepackage{amsmath,amssymb,amsfonts}
\usepackage{algorithmic}
\usepackage{comment}
\usepackage{graphicx}
\usepackage{textcomp}
\usepackage{subcaption}
\usepackage{url}
\usepackage{multirow}
\usepackage{xcolor}
\usepackage{pifont}
\usepackage{booktabs}
\def\BibTeX{{\rm B\kern-.05em{\sc i\kern-.025em b}\kern-.08em
    T\kern-.1667em\lower.7ex\hbox{E}\kern-.125emX}}
\begin{document}

\title{Automated Glaucoma Report Generation via Dual-Attention Semantic Parallel-LSTM and Multimodal Clinical Data Integration\\
}

\author{\IEEEauthorblockN{Cheng Huang$^{1,2}$, Weizheng Xie$^{1}$, Zeyu Han$^{1}$, Tsengdar Lee$^{3,\dag}$\\ Karanjit Kooner$^{2}$, Jui-Ka Wang$^{2}$, Ning Zhang$^{4}$ and Jia Zhang$^{1,\dag}$}
\IEEEauthorblockA{\textit{$^{1}$Southern Methodist University, $^{2}$University of Texas Southwestern Medical Center} \\
\textit{$^{3}$National Aeronautics and Space Administration, $^{4}$Northeastern University}\\
$^{\dag}$Corresponding Author; E-mail: jiazhang@smu.edu}
}

\maketitle

\begin{abstract}
Generative AI for automated glaucoma diagnostic report generation faces two predominant challenges: content redundancy in narrative outputs and inadequate highlighting of pathologically significant features including optic disc cupping, retinal nerve fiber layer defects, and visual field abnormalities. These limitations primarily stem from current multimodal architectures' insufficient capacity to extract discriminative structural-textural patterns from fundus imaging data while maintaining precise semantic alignment with domain-specific terminology in comprehensive clinical reports. To overcome these constraints, we present the Dual-Attention Semantic Parallel-LSTM Network (DA-SPL), an advanced multimodal generation framework that synergistically processes both fundus imaging and supplementary visual inputs. DA-SPL employs an Encoder-Decoder structure augmented with the novel joint dual-attention mechanism in the encoder for cross-modal feature refinement, the parallelized LSTM decoder architecture for enhanced temporal-semantic consistency, and the specialized label enhancement module for accurate disease-relevant term generation. Rigorous evaluation on standard glaucoma datasets demonstrates DA-SPL's consistent superiority over state-of-the-art models across quantitative metrics. DA-SPL exhibits exceptional capability in extracting subtle pathological indicators from multimodal inputs while generating diagnostically precise reports that exhibit strong concordance with clinical expert annotations.
\end{abstract}

\begin{IEEEkeywords}
Fundus Imaging, Image Caption, Medical Report Generation, Attention Mechanism, Multimodal Model, Generative AI
\end{IEEEkeywords}

\section{Introduction}

Glaucoma, a progressive optic neuropathy, remains one of the leading causes of irreversible blindness worldwide \cite{fairclip,fairdomain,fairseg}. The disease is characterized by gradual loss of retinal ganglion cells, structural deformation of the optic nerve head, and corresponding visual field defects \cite{bg-1,lag2}. Because early glaucomatous damage often develops silently and progresses without noticeable symptoms until late stages, a substantial proportion of patients remain undiagnosed or receive treatment only after irreversible vision loss has occurred \cite{bg-2,lag}. Consequently, early detection, precise staging, and continuous monitoring are critical for preventing visual impairment and guiding clinical management.

Recent advances in artificial intelligence (AI) have greatly facilitated the automated analysis of ocular images, particularly color fundus photographs and optical coherence tomography (OCT) scans, for glaucoma screening \cite{fairclip}, segmentation of retinal structures \cite{fairseg}, and progression prediction \cite{fairvision}. Despite these achievements, most existing approaches are primarily computer vision oriented, focusing on pattern recognition and pixel-level inference tasks \cite{SCES,STARE,origa,rimone_dl,icct,9478347}. Such methods often operate as black-box systems that yield quantitative indices or segmentation maps but fail to convey clinically interpretable reasoning or integrate domain-specific ophthalmic knowledge into diagnostic narratives. This lack of semantic alignment with expert reasoning limits clinical adoption, as ophthalmologists depend not only on image features but also on contextual cues such as patient history, anatomical variations, and inter-eye asymmetry.

Therefore, there is an urgent need for AI frameworks that go beyond visual analysis toward generating diagnostically coherent and knowledge-driven reports. Such systems should bridge the gap between imaging biomarkers and textual medical interpretation, providing structured outputs that mirror the logical reasoning of human experts. Developing these multimodal, knowledge-integrated models would represent a crucial step toward trustworthy, explainable, and clinically actionable glaucoma assessment.

To bridge this critical gap, we propose the \textbf{D}ual-\textbf{A}ttention \textbf{S}emantic \textbf{P}arallel-\textbf{L}STM Network (\textbf{DA-SPL}), a novel framework that synergizes visual feature extraction with semantic understanding to transform image-based analysis into clinically interpretable diagnostic reports. By integrating advanced computer vision and natural language processing techniques, DA-SPL generates comprehensive reports that maintain both diagnostic accuracy and clinical relevance.

The principal contributions of this work are three-fold:

\begin{itemize}
    \item We introduce DA-SPL, an innovative glaucoma report generation framework that effectively combines multimodal visual features with semantic understanding through a dual-attention encoder and parallel LSTM decoder architecture, enhanced by a specialized label refinement module for improved diagnostic precision.
    
    \item We propose a novel adaptive weight balancing mechanism within DA-SPL that optimizes feature representation learning by dynamically adjusting to the intrinsic statistical properties of the dataset, resulting in more accurate and clinically relevant report generation.
    
    \item Through extensive experimentation on benchmark glaucoma datasets, we demonstrate the superior performance of our proposed architecture and provide comprehensive ablation studies validating the contribution of each component.
\end{itemize}

\section{Related Work}

\subsection{AI For Glaucoma}

Over the past decade, a variety of publicly available datasets have been introduced to accelerate research at the intersection of artificial intelligence and glaucoma diagnosis. Most existing resources primarily focus on computer vision tasks using retinal fundus or OCT images. For instance, the LAG dataset \cite{lag} contains 5,824 fundus images, serving as a benchmark for automated glaucoma detection. Similarly, OCTA-500 \cite{octa500} provides optical coherence tomography angiography (OCTA) images from 500 subjects, supporting cross-modality vascular analysis. Large-scale resources such as Harvard-GDP \cite{gdp} and Harvard-GF \cite{gf} include 21,059 and 3,300 retinal samples, respectively, enabling population-level investigations into glaucomatous features. Several mid-sized datasets have contributed to optic disc and cup segmentation, including FairSeg \cite{fairseg}, FairDomain \cite{fairdomain}, and classical collections such as Drishti-GS \cite{drishti}, ORIGA \cite{origa}, and RIM-ONE \cite{rimone_dl}. Foundational retinal datasets such as DRIVE \cite{ridge}, SCES \cite{SCES}, and STARE \cite{STARE} have been widely reused for vessel segmentation and general fundus analysis, while more recent efforts including MURED \cite{mured}, ROSE \cite{rose}, and RetinaMix \cite{RetinaMix} have emphasized diverse imaging modalities and fair demographic representation.

As shown in these resources, current AI research in glaucoma predominantly focuses on computer vision tasks, employing convolutional neural networks (CNNs) \cite{cnn,lag,lag2} and their derivatives such as U-Net \cite{unet} for applications including pathological feature detection \cite{fairclip,fairvision,glagan}, optic disc segmentation \cite{fairseg}, and three-dimensional structural reconstruction \cite{octa500}. These approaches have demonstrated strong capability in extracting low-level visual representations such as retinal layer thickness patterns \cite{rose,drishti}, yet they remain limited in associating these imaging findings with clinically interpretable biomarkers, including intraocular pressure dynamics and optic nerve head deformation \cite{bg-2}. Although the combination of multimodal data and clinical reasoning represents a promising frontier, the domain of automated diagnostic report generation, particularly approaches integrating natural language processing corpora with vision-language or large language models \cite{gpt4o,llama3,r1,v3,aensi,svaml}, remains largely unexplored within glaucoma-oriented artificial intelligence research. This research gap highlights the necessity for next-generation frameworks capable of unifying imaging biomarkers and textual medical reasoning to produce comprehensive, interpretable, and clinically aligned diagnostic outputs.

\subsection{Medical Report Generation for Glaucoma}

Automated medical report generation has recently emerged as a transformative paradigm for converting diagnostic imaging data into structured clinical narratives. This approach offers significant potential for reducing physician workload, enhancing reporting consistency, and improving interpretability in clinical decision-making \cite{IFE}. Substantial progress has been achieved in radiological imaging, particularly in the automatic interpretation of chest X-rays \cite{llama3}. However, comparable advancements in ophthalmology, and glaucoma in particular, remain considerably limited. Existing studies in other medical domains often employ hybrid architectures that combine computer vision backbones such as convolutional neural networks (CNNs) with sequential language models including recurrent neural networks (RNNs) or long short-term memory (LSTM) models \cite{cnn-rnn,cnn-lstm}. More recent approaches have extended this framework by introducing vision-language models that align visual representations with textual semantics to generate coherent diagnostic reports.

Applying such methodologies to glaucoma analysis introduces several unique challenges. First, publicly available datasets that pair fundus or OCT images with expert-annotated clinical reports are scarce, restricting the development of data-driven multimodal systems \cite{glaboost,glalstm}. Second, most existing architectures overlook critical glaucoma-specific biomarkers, such as cup-to-disc ratio, retinal nerve fiber layer (RNFL) thickness, and peripapillary vessel density, which are essential for accurate disease characterization. Third, glaucoma is inherently a progressive disorder that evolves slowly over time, yet current report generation frameworks rarely incorporate temporal modeling or longitudinal progression patterns. These limitations collectively underscore the necessity for specialized multimodal frameworks capable of integrating imaging biomarkers, clinical domain knowledge, and temporal dynamics to generate accurate and clinically meaningful glaucoma diagnostic reports.

\begin{figure*}
    \centering
    \includegraphics[width=1.0\linewidth]{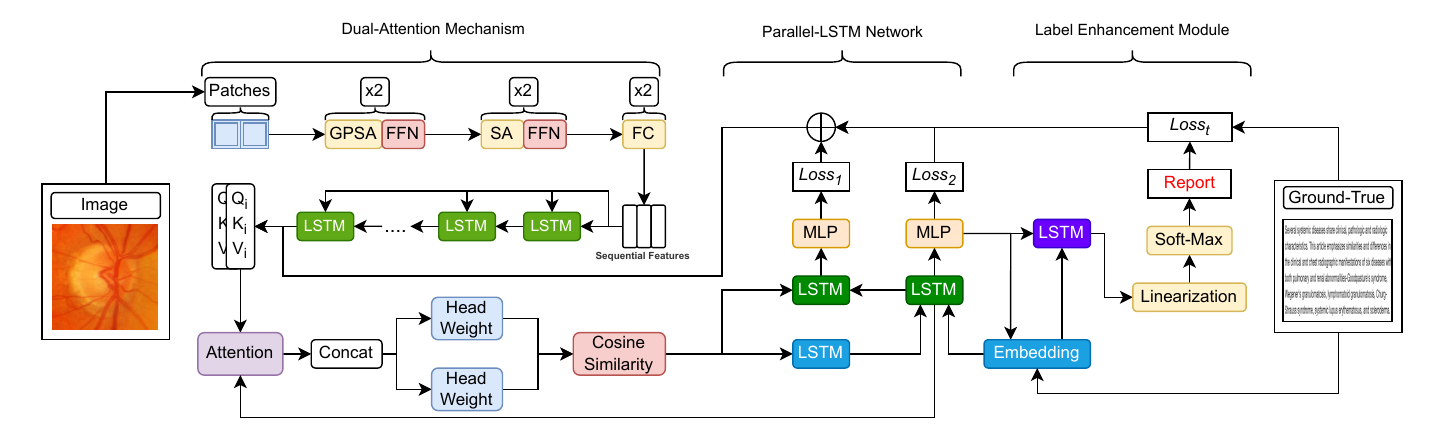}
    \caption{The Architecture of DA-SPL}
    \label{structure}
\end{figure*}

\section{Methodological Framework}

As illustrated in Fig.~\ref{structure}, DA-SPL follows an encoder-decoder paradigm with three key components: 
\begin{itemize}
    \item \textit{Dual-Attention Mechanism (DAM)}
    \item \textit{Parallel-LSTM Network (PLN)}
    \item \textit{Label Enhancement Module (LEM)}
\end{itemize}
Each component is described in detail below, followed by the formulation of our optimization objective.

\subsection{Dual-Attention Mechanism}

Our encoder begins by partitioning the input fundus image into non-overlapping patches, which are subsequently processed through gated positional self-attention (GPSA) layers \cite{vit,convit}. These layers incorporate positional embeddings to maintain spatial relationships within the retinal structure. The feedforward network (FFN) component \cite{att} employs a two-layer architecture with Gaussian Error Linear Unit (GeLU) activation \cite{gelu} for nonlinear transformation. Following the final GPSA layer, we append a classification token and project the features through a fully connected layer to obtain 1D sequential representations \cite{cnn,convit,resnet}. These features are then processed by the initial LSTM unit \cite{lstm,IFE}, coupled with an attention mechanism for iterative refinement before transformer-based processing \cite{att}.

The conventional multi-head attention mechanism is enhanced through our dual-weighting approach. The standard multi-head attention computation is given by:
\begin{gather}
att^{i}{a} = softmax\left(\frac{QK^{T}}{\sqrt{d{k}}}\right)V \\
att_{a} = Concat(att^{1}_{a}, att^{2}_{a},..., att^{N}_{a})W
\label{eq}
\end{gather}
where $N$ is the number of heads, and $d_k$ the scaling factor.

Fot the first weight: learnable head weights $w_a$ are updated each iteration via
\begin{equation}
w_{a(i)} = softmax(w_{a(i-1)})N
\label{eq2}
\end{equation}
and applied to the attention output:
\begin{equation}
att_{wa(i)} = w_{a(i)} * att_{a(i)}
\label{eq3}
\end{equation}
Initial weights are uniform; training assigns higher values to heads with greater task relevance.

Fot the second weight, to enhance salient heads, we compute cosine similarity with the most important head in the batch:
\begin{equation}
cos(i)^{j} = cos\left(att^{j}_{a(i-1)}, att^{base}_{a(i-1)}\right)
\label{eq4}
\end{equation}
Compared with the single-weight configuration, as demonstrated in TABLE~\ref{t1}, the dual-weight implementation exhibits a twofold increase in parameter magnitude. This enhanced weighting scheme within the attention mechanism enables more robust feature emphasis, facilitates multi-scale information aggregation, and yields superior representational capacity while maintaining model generalizability.

\begin{table}[h]
\caption{Comparison of Single Weight \& Double Weight}
\label{t1}
\centering
\tiny
\scalebox{0.95}{
\begin{tabular}{l|l|cccccccc}
\hline
\multirow{2}{*}{\textbf{Fold}} & \multirow{2}{*}{\textbf{Weight}} & \multicolumn{8}{c}{\textbf{Head}}  \\
\cline{3-10}&  &  \textbf{1} &  \textbf{2} &  \textbf{3} &  \textbf{4} &  \textbf{5} &  \textbf{6} &  \textbf{7} &  \textbf{8}\\
\hline
\multirow{2}{*}{1} 
& single & 2.737& \textcolor{blue}{7.981}  & 2.737& 2.737 & 2.737 & 2.737 & 2.737& 2.737 \\  
& double & 5.417& \textcolor{blue}{14.964} & 5.444& 5.419& 5.508& 5.501& 5.407& 5.425\\
\hline
\multirow{2}{*}{2}
& single & \textcolor{blue}{7.981}& 2.737& 2.737& 2.737& 2.737& 2.737& 2.737& 2.737 \\
& double & \textcolor{blue}{14.964} & 5.392& 5.476& 5.599& 5.418& 5.490& 5.417& 5.458\\
\hline
\multirow{2}{*}{3}
& single & 2.737& 2.737& 2.737& 2.737& 2.737& 2.737& \textcolor{blue}{7.981}& 2.737 \\
& double & 5.511& 5.473& 5.452& 5.425& 5.523& 5.444& \textcolor{blue}{14.964} & 5.512\\
\hline
\multirow{2}{*}{4}
& single & 2.737& 2.737& 2.737& 2.737& 2.737& \textcolor{blue}{7.981} & 2.737& 2.737 \\
& double & 5.425& 5.376& 5.455& 5.449& 5.392& \textcolor{blue}{14.964} & 5.388& 5.338\\
\hline
\multirow{2}{*}{5}
& single & \textcolor{blue}{7.981} & 2.737& 2.737& 2.737& 2.737& 2.737& 2.737& 2.737 \\
& double & \textcolor{blue}{14.964} & 5.436& 5.573& 5.399& 5.415& 5.466& 5.484& 5.506\\
\hline
\multirow{2}{*}{6}
& single & 2.737& 2.737& 2.737& 2.737& 2.737& 2.737& 2.737& \textcolor{blue}{7.981} \\
& double & 5.526& 5.575& 5.525& 5.506& 5.570& 5.486& 5.558& \textcolor{blue}{14.964} \\
\hline
\multirow{2}{*}{7}
& single & 2.737& 2.737& 2.737& 2.737& 2.737& 2.737& \textcolor{blue}{7.981}& 2.737 \\
& double & 5.567& 5.519& 5.489& 5.431& 5.531& 5.524& \textcolor{blue}{14.964} & 5.486 \\
\hline
\multirow{2}{*}{8}
& single & 2.737& 2.737& 2.737& 2.737& \textcolor{blue}{7.981}& 2.737& 2.737& 2.737 \\
& double & 5.464& 5.474& 5.495& 5.514& \textcolor{blue}{14.964} & 5.542& 5.480& 5.495\\
\hline
\multirow{2}{*}{9}
& single & 2.737& 2.737& 2.737& \textcolor{blue}{7.981}& 2.737& 2.737& 2.737& 2.737 \\
& double & 5.422& 5.662& 5.387& \textcolor{blue}{14.964} & 5.342& 5.435& 5.445& 5.390\\
\hline
\multirow{2}{*}{10}
& single & 2.737& 2.737& \textcolor{blue}{7.981}& 2.737& 2.737& 2.737& 2.737& 2.737 \\
& double & 5.591& 5.502& \textcolor{blue}{14.964} & 5.530& 5.523& 5.517& 5.506& 5.592\\
\hline
\multicolumn{10}{c}{\textcolor{blue}{\textbf{Blue}} represents a number without extra units.} \\
\multicolumn{10}{c}{\textcolor{black}{\textbf{Black}} means there is a unit ($\times10^{-3}$) after the number.} \\
\hline
\end{tabular}}
\end{table}

Averaging over the batch yields:
\begin{equation}
w_{cos(i)}^{j} = \frac{\sum_{k\in i} cos(i)_{k}^{j}}{N}
\label{eq5}
\end{equation}

For the combination of Dual-Weight, We define
\begin{equation}
\label{eq7}
\beta = \sqrt[n]{\prod^{n}_{1} \lvert w_{cos(i)} \rvert}
\end{equation}
and adjust the cosine weight by
\begin{equation}
w^{j}_{dwa(i)} = \beta - w_{cos(i)}
\label{eq6}
\end{equation}
This mechanism selectively amplifies attention heads corresponding to pathological samples while attenuating those associated with normal anatomical features. The ReLU-based weight rectification eliminates negative activations, thereby enhancing computational efficiency and sharpening the model's focus on clinically relevant pathological indicators - effectively mimicking the selective attention processes observed in human diagnostic reasoning.

\subsection{Parallel-LSTM Network}

The Parallel-LSTM Network consists of three LSTMs but two of them are the same. LSTM (blue, as shown in Fig.~\ref{structure}) is just used to encode features from Dual-Attention Mechanism, as expressed in Eq. \ref{eq8}.
\begin{equation}
\label{eq8}
h^{1}_{t} = LSTM(T^{1}_{t},w^{j}_{dwa(i)}*att^{j}_{a(i)})
\end{equation}
where $T^{1}_{t}$ denotes the word embedding of transformer generated. 
When the first LSTM (dark green) decode the first time, the inputs are $T_{t}^{1}$, $h_{t}^{1}$, the word embedding $W_{e}^{2}x_{t}$ and the hidden state of LSTM, $h^{2}_{t-1}$, as shown in Eq. \ref{eq9}.
\begin{equation}
\label{eq9}
h_{t}^{2}= LSTM(T_{t}^{1}, h_{t}^{1}, W_{e}^{2}x_{t}, h_{t-1}^{2})
\end{equation}
Then the probability of next word can be predicted by a MLP, as appeared in Eq. \ref{eq10}:
\begin{equation}
\label{eq10}
MLP_{t+1}^{1}= softmax(W_{FC1}*h_{t}^{2})
\end{equation}
Where $W_{FC1}$ is the fully connected layer of MLP. So does the sencond LSTM (dark green), as shown in Eq. \ref{eq11}.
\begin{gather}
h_{t}^{3}= LSTM(T_{t}^{2}, h_{t}^{2}, W_{e}^{2}x_{t}) \\
MLP_{t+1}^{1}= softmax(W_{FC2}*h_{t}^{3})
\label{eq11}
\end{gather}
After that, the text features generated by the two MLPs are used to calculate the loss functions and use it in the later stage.

\subsection{Label Enhancement Module}

Medical imaging datasets typically comprise disease-specific annotations and clinical descriptors that are critical for multi-label classification tasks. However, limited dataset scales coupled with extensive label spaces often compromise classification precision, subsequently degrading the quality of generated diagnostic reports. To address this, we conceptualize diagnostic labels as report "keywords" and propose a dedicated quality refinement module in DA-SPL that performs post-generation report optimization. Departing from conventional approaches, our novel multi-label classifier operates on the generated report text rather than direct image input. The architecture integrates a category prediction LSTM (illustrated in purple) with a subsequent classification layer, processing the embedded report representation $RP_{emb}$ through the following computational pipeline:

The module first processes $RP_{emb}$ through the category prediction LSTM to generate initial label predictions $Label$, as formalized in Eq.\ref{Label}, 
\begin{equation}
\label{Label}
Label=softmax(W_{t}*LSTM(RP_{emb})+b_{t})
\end{equation}
where $W_t$ and $b_t$ are trainable parameters for flattening the LSTM output. A mapping layer aligns $Label$ with the classification dimensionality, and a softmax yields the probability distribution over all possible tags. This label-focused quality enhancement module refines report accuracy by aligning generated content with diagnostic categories, enabling more precise and clinically relevant medical image report generation.

\subsection{Loss Function}

As shown in Eq. \ref{loss}, $loss$ is the final loss of model, $loss1$ ($loss2$) is the cross-entropy loss between reports generated by first (second) LSTM  in the Parallel-LSTM Network and true captions. The second LSTM (dark green) generation is confined by an adjustable parameter $\lambda\in(0,1]$. From the beginning, we set $\lambda$ to 0.5 at first.
\begin{equation}
\label{loss}
l o s s=l o s s_{1}+\lambda*l o s s_{2}+\alpha*l o s s_{t}
\end{equation}
Since $loss_{t}$ is merely one-tenth of $loss_{1}$ and $loss_{2}$ statistically, $loss_{t}$ needs to be amplified to balance. Given that $loss_{t}$ does not directly reflect to reports quality like $loss_{1}$ and $loss_{2}$, we set its coefficient $\alpha\in[1,10)$ at 5. Among them, $loss_{t}$\footnote{MultiLabel- SoftMarginLoss: https://pytorch.org/docs} can be expressed as Eq. \ref{loss-t}:
\begin{align}
\begin{aligned}
\label{loss-t}
  loss_{t} & = -\frac{1}{C}*\sum label_{i} * log(sig(Label_{i})) \\ & + (1-label_{i}) * log(1-sig(Label_{i}))
\end{aligned}
\end{align}
Where $label$ represents the true label, $i\in\{0,...,n-1\}$, $label_{i}\in\{0,1\}$; $n$ is the number of labels types. 
$loss_{t}$ is added in backward propagation of training process for Label Enhancement Module.

\section{Dataset and Implementation}

\subsection{Dataset}

As shown in TABLE~\ref{data-public}, the amount of data has been reduced significantly. For its preprocessing to  features like “optic\_disc\_size” which has specific value (feature, also called factor, structured data), we will refer to the judgment criteria for glaucoma and give ``True" or ``False". “neuroretinal\_rim” is just the description (corpus). 
“glaucoma\_risk\_assessment” and “confidence\_level” are determined by human subjective judgment, akin to labels, indicating whether the case is classified as glaucoma and providing an individual's evaluation of the judgment. Dataset with its images is available through this link\footnote{\url{https://huggingface.co/datasets/AswanthCManoj/glaucoma_diagnosis_json_analysis}}.

\begin{table}[h]
\caption{One Sample from Public Dataset}
\begin{center}
\begin{tabular}{l|l}
\hline
\textbf{Fundus\_Features} & \textbf{Ground True} \\
\hline
optic\_disc\_size  & large \\
cup\_to\_disc\_ratio  & 0.8 \\
isnt\_rule\_followed  & false \\
rim\_pallor  & true \\
rim\_color  & pale \\
bayoneting  & true \\
sharp\_edge  & true \\
laminar\_dot\_sign  & true \\
notching  & true \\
rim\_thinning  &  true \\
additional\_observations  & null  \\
neuroretinal\_rim  & Description  \\
\hline
glaucoma\_risk\_assessment  &  high risk\\
confidence\_level  & 0.9 \\
\hline
\end{tabular}
\label{data-public}
\end{center}
\end{table}

\subsection{Implementation}

\subsubsection{Hyperparameter}
we utilizes a ConViT \cite{convit}, pretrained on ImageNet \cite{imagenet_cvpr09}, with the final classification layer removed to extract 512-dimensional visual feature representations. Both word embeddings and LSTM hidden states are configured with 512-dimensional vectors \cite{lstm,IFE}. Model optimization employs the ADAM optimizer \cite{adamw} with weight decay. The baseline architecture combines ConViT with standard multi-head attention \cite{att}, trained with an initial learning rate of 0.0004. Through empirical validation, we set the trade-off parameter $\lambda=0.5$ for optimal performance. Text generation utilizes beam search with width 5, and automated metric computation is performed using the standard image captioning evaluation toolkit.

\subsubsection{Evaluation Metrics}
we assess model performance using three established automatic evaluation metrics for natural language generation: BLEU \cite{bleu}, CIDEr \cite{cider}, and ROUGE-L \cite{rouge}. The BLEU metric (B-n, where n denotes n-gram order) calculates modified n-gram precision with a brevity penalty to penalize truncated outputs. CIDEr (CID) employs TF-IDF weighted n-gram matching to emphasize clinically relevant terminology. ROUGE-L (ROU) evaluates the longest common subsequence to measure content coverage. This comprehensive metric suite evaluates report fluency, clinical relevance, and semantic alignment with reference standards. In result tables, the top-performing values are highlighted in bold, while suboptimal results are underlined.

\subsubsection{Computational Environment}
the experimental platform consists of a GPU cluster with the following specifications: 2 $\times$ NVIDIA Tesla V100 GPUs (32GB VRAM each), 64GB system RAM, 8-core CPUs per compute node, and a total of 6 interconnected nodes. All experiments were conducted using PyTorch with mixed-precision training enabled.

\section{Experimental Results}

\subsection{Experimental Configuration}
\subsubsection{Comparative Model} 
we benchmark against five state-of-the-art medical report generation architectures: CNN-RNN \cite{cnn-rnn}, CNN-LSTM \cite{cnn-lstm}, CaptionNet \cite{IFE}, AoANet \cite{aoa}, and JE-TriNet \cite{joint}. All comparative models were hyperparameter-optimized using Optuna\footnote{\url{https://optuna.org/}} with Bayesian optimization. Each model underwent rigorous training for 200 epochs, with performance evaluation conducted through ten-fold cross-validation to ensure statistical reliability and generalization capability. The validation protocol maintains identical data splits across all compared methods.

\subsubsection{Large Language Model}
to enhance medical report generation, we incorporate both open-source (LlaMA family \cite{llama,llama2,llama3}) and proprietary (GPT series \cite{gpt4o,o3}) large language models, fine-tuned for the downstream diagnostic task. The LlaMA models (3.1-70B and 3.1-405B) were deployed via the LlaMA-API platform, while GPT-4O and GPT-O1 were accessed through their official APIs. For visual feature extraction, we employ a CNN backbone (ResNet-152 \cite{cnn,resnet}), whose final fully-connected layers are projected into the LLM's embedding space. The fine-tuning process, detailed in TABLE~\ref{tab:full-finetune}, involves: (1) augmenting the LLM with a task-specific vision adapter head, (2) joint optimization of both components using AdamW \cite{adamw} with cosine learning rate decay (10\% warm-up), and (3) cross-entropy loss minimization\footnote{\url{https://docs.pytorch.org/docs/stable/generated/torch.nn.CrossEntropyLoss}}. This approach enables the LLM to develop task-specific representations while preserving its generative capabilities.

\begin{table}[!ht]
\centering
\caption{Fine-tuning with Vision Encoder Head}
\label{tab:full-finetune}
\scalebox{0.9}{
\begin{tabular}{l|l|cccccc}
\hline
\textbf{LLM} & \textbf{Version}  & \textbf{DR}  & \textbf{LR} & \textbf{BS} & \textbf{MSL} & \textbf{Epoch}   & \textbf{GC}  \\
\hline
\multirow{4}{*}{GPT} 
  & 4o         & $5\!\times\!10^{-5}$  & $5\!\times\!10^{-5}$ & 8 & 512 & 50  & 1.0  \\
  & o1         & $5\!\times\!10^{-5}$  & $5\!\times\!10^{-5}$ & 8 & 512 & 50  & 1.0  \\
  & o1-mini         & $7\!\times\!10^{-5}$  & $7\!\times\!10^{-5}$ & 8 & 512 & 50  & 1.0  \\
  & 4.1         & $5\!\times\!10^{-5}$  & $5\!\times\!10^{-5}$ & 8 & 512 & 50  & 1.0  \\
\hline

\multirow{3}{*}{LlaMA}  
  & 3.1-405B   & $5\!\times\!10^{-5}$  & $5\!\times\!10^{-5}$ & 8 & 512 & 50  & 1.0  \\
  & 3.1-70B    & $5\!\times\!10^{-5}$  & $5\!\times\!10^{-5}$ & 8 & 512 & 50  & 1.0  \\
  & 3.1-8B    & $7\!\times\!10^{-5}$  & $7\!\times\!10^{-5}$ & 8 & 512 & 50  & 1.0  \\
\hline
\end{tabular}}
\end{table}

\noindent\textbf{Note:} Dropout Rate = DR; Learning Rate = LR; Batch Size = BS; Max Sequence Length = MSL; Gradient Clipping = GC and Gradient Clipping = GC.

\subsection{Comparative Experiment}

Table \ref{ce-1} reports the performance of all models on the glaucoma report generation task. While JE-TriNet \cite{joint} yields the strongest baseline results and GPT-4o \cite{gpt4o} leads among LLMs, DA-SPL achieves the highest overall performance, with notable gains in BLEU-1, BLEU-2, BLEU-4, ROUGE-L, and CIDEr, and a competitive BLEU-3, underscoring its effectiveness in producing accurate and semantically comprehensive reports.

\begin{table}[h]
    \centering
    \caption{Comparative Experiment Among Models ($\times$100\%)}
    \label{ce-1}
    \scalebox{0.87}{
    \begin{tabular}{l|l|cccccc}
    \hline
    \multicolumn{2}{l|}{\textbf{Baseline Model}}   & \textbf{B-1} &  \textbf{B-2} & \textbf{B-3} & \textbf{B4}  & \textbf{ROU} & \textbf{CID} \\
    \hline
    CNN-RNN     & Optuna & 42.13 & 26.34 & 17.25 & 11.28 & 36.27 & 32.41 \\
    CNN-LSTM    & Optuna & 46.52 & 29.18 & 19.46 & 13.47 & 38.49 & 37.95 \\
    CaptionNet  & Optuna & 49.27 & 31.56 & 21.38 & 14.52 & 40.23 & 45.28 \\
    AoANet      & Optuna & 52.79 & 33.82 & 23.74 & 16.25 & 43.56 & 56.31 \\
    JE-TriNet   & Optuna & \underline{54.38} & \underline{36.17} & \textbf{25.49} & \underline{17.93} & \underline{44.87} & \underline{61.84} \\
    \hline
    \textbf{LLMs} & \textbf{Encoder} & \multicolumn{6}{c}{\textbf{-}}  \\
    \hline
    LlaMa3.1-8B      & CNN  & 50.28 & 33.69 & 22.95 & 16.24 & 42.85 & 57.86 \\
    LlaMa3.1-70B     & CNN  & 51.72 & 34.98 & 24.07 & 17.15 & 43.79 & 59.64 \\
    LlaMa3.1-405B    & CNN  & 52.36 & 35.51 & 24.59 & 17.53 & 44.12 & 60.48 \\
    GPT-4o           & CNN  & 52.94 & 35.98 & 24.97 & 17.86 & 44.45 & 61.23 \\
    GPT-o1           & CNN  & 52.05 & 35.27 & 24.31 & 17.39 & 43.92 & 60.11 \\
    GPT-o1-mini      & CNN  & 50.63 & 33.98 & 23.14 & 16.42 & 43.06 & 58.37 \\
    GPT-4.1          & CNN  & 52.68 & 35.79 & 24.78 & 17.72 & 44.31 & 60.94 \\

    \hline
    \multicolumn{2}{l|}{DA-SPL} & \textbf{56.97} & \textbf{36.71}  & \underline{25.14} & \textbf{18.74} & \textbf{45.08} & \textbf{62.44} \\
    \hline 
    \end{tabular}}
\end{table}

\subsection{Ablation Experiment}

\subsubsection{Proposed Component}

ablation experiments were conducted to evaluate the contribution of each proposed component. As shown in TABLE~\ref{ab-1}, performance improves steadily as modules are added, with LEM providing the highest single-module gain, followed closely by PLN and DAM. Applying hyperparameter tuning via Optuna further enhances results across all configurations. DA-SPL with all modules and tuning achieves the best overall performance, exceeding the untuned configuration by approximately two percentage points across the evaluation metrics.

\begin{table}[h]
    \centering
    \caption{Ablation Study on Proposed Module ($\times$100\%)}
    \label{ab-1}
    \scalebox{0.97}{
    \begin{tabular}{l|cccccc}
    \hline
     \textbf{\textbf{Module}}  & \textbf{B-1} &  \textbf{B-2} & \textbf{B-3} & \textbf{B4}  & \textbf{ROU} & \textbf{CID} \\
    \hline
+ DAM             & 10.42 & 6.85 & 4.02 & 2.91 & 8.64 & 14.73 \\
    + DAM (Optuna)    & 24.36 & 15.28 & 9.83 & 7.12 & 20.17 & 28.45 \\
    \hline
    + PLN             & 31.59 & 20.47 & 13.26 & 9.54 & 26.83 & 37.68 \\
    + PLN (Optuna)    & 39.74 & 25.83 & 16.87 & 12.16 & 32.74 & 45.92 \\
    \hline
    + LEM             & 46.12 & 30.21 & 19.83 & 14.23 & 37.89 & 53.15 \\
    + LEM (Optuna)    & 52.35 & 34.68 & 23.15 & 16.72 & 42.36 & 58.79 \\
    \hline
    + all             & \underline{55.12} & \underline{35.89} & \underline{24.52} & \underline{18.12} & \underline{44.63} & \underline{61.02} \\
    + all (Optuna)  & \textbf{56.97} & \textbf{36.71} & \textbf{25.14} & \textbf{18.74} & \textbf{45.08} & \textbf{62.44} \\
    \hline 
    \end{tabular}}
\end{table}

\subsubsection{Backbone \& Weight}

TABLE~\ref{ab-2} presents the ablation results on backbone selection and attention weight strategy. Within the DAM setting, replacing the standard single-weight attention with the dual-weight variant yields slight but consistent improvements across most metrics for the ViT backbone. When comparing backbones, ConViT outperforms ViT under the same attention configuration. Furthermore, for the ConViT backbone, adopting the dual-weight design in the proposed DA-SPL leads to the highest performance across all metrics, confirming the effectiveness of both the backbone choice and the dual-weight attention mechanism.

\begin{table}[h]
    \centering
    \caption{Ablation Study on BackBone \& Weight ($\times$100\%)}
    \label{ab-2}
    \scalebox{0.97}{
    \begin{tabular}{l|cccccc}
    \hline
     \textbf{DAM Backbone }  & \textbf{B-1} &  \textbf{B-2} & \textbf{B-3} & \textbf{B4}  & \textbf{ROU} & \textbf{CID} \\
     \hline
     ViT (DAM) & 54.82 & 35.12 & 23.64 & 17.25 & 43.72 & 60.18 \\
     ViT (Dual-weight) & 54.95 & 35.27 & 23.79 & 17.39  & \underline{44.25} & 60.36 \\
    \hline
     \textbf{Weight (ConViT)}  & \textbf{B-1} &  \textbf{B-2} & \textbf{B-3} & \textbf{B4}  & \textbf{ROU} & \textbf{CID} \\
     \hline
    Single-weight  & \underline{55.96} & \underline{35.84} & \underline{24.21} & \underline{17.88}& 43.85 & \underline{61.07} \\
    DA-SPL  & \textbf{56.97} & \textbf{36.71} & \textbf{25.14} & \textbf{18.74} & \textbf{45.08} & \textbf{62.44} \\
    \hline 
    \end{tabular}}
\end{table}

\begin{figure*}[t]
  \centering
  \begin{subfigure}[t]{0.48\textwidth}
    \centering
    \includegraphics[width=\linewidth]{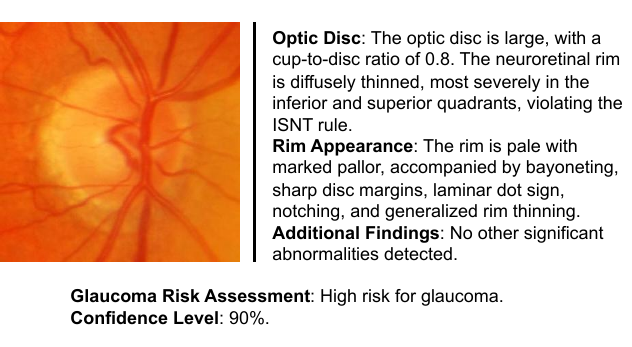}
    \caption{Glaucoma Report}
    \label{fig:r1}
  \end{subfigure}\hfill
  \begin{subfigure}[t]{0.48\textwidth}
    \centering
    \includegraphics[width=\linewidth]{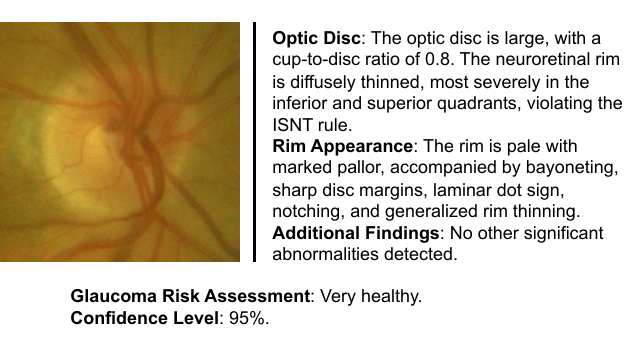}
    \caption{Nomal Eye Report}
    \label{fig:r2}
  \end{subfigure}
  \caption{Medical Report Generation of DA-SPL on Glaucoma \& Nomal Eye}
  \label{fig:r}
\end{figure*}

\subsubsection{Loss Function} 
we set $\alpha=5$ for optimal performance, as determined by an ablation in which we varied $\alpha \in [1,10]$. As shown in TABLE~\ref{ab-3}, ablation experiments varying $\alpha$ from 1 to 10 reveal a clear performance peak at $\alpha=5$, where most key metrics reach their highest or near-highest values. This suggests that $\alpha=5$ provides an optimal balance between the relative contributions in the loss function, enabling the model to effectively learn salient features without overemphasizing or underweighting specific components.

\begin{table}[h]
    \centering
    \caption{Impact of $\alpha$ on Model Performance ($\times$100\%)}
    \label{ab-3}
    \scalebox{1.0}{
    \begin{tabular}{l|cccccc}
    \hline
     \textbf{$\alpha$}  & \textbf{B-1} &  \textbf{B-2} & \textbf{B-3} & \textbf{B4}  & \textbf{ROU} & \textbf{CID} \\
    \hline
1  & 38.42 & 25.31 & 16.85 & 11.94 & 29.78 & 28.63 \\
2  & 42.57 & 28.46 & 18.92 & 13.02 & 33.14 & 31.25 \\
3  & 51.86 & 33.47 & 22.75 & 16.59 & 41.03 & 55.26 \\
4  & 55.62 & 35.94 & 24.89 & 18.33 & \textbf{45.08} & 61.08 \\
5  & \textbf{56.97} & \underline{36.71} & \textbf{25.14} & \textbf{18.74} & \textbf{45.08} & \underline{62.44} \\
6  & 56.21 & \textbf{36.98} & 24.96 & 18.59 & \underline{44.95} & 61.92 \\
7  & 54.18 & 35.26 & 24.37 & 18.02 & 44.01 & 60.79 \\
8  & 40.25 & 27.14 & 18.05 & 12.41 & 31.92 & 29.87 \\
9  & 36.83 & 24.95 & 16.47 & 11.08 & 28.65 & 27.53 \\
10 & 33.94 & 22.83 & 15.26 & 10.27 & 26.48 & 25.39 \\
    \hline 
    \end{tabular}}
\end{table}

\subsubsection{Input Modality}

Table~\ref{ab-4} reports the impact of different input modality combinations on DA-SPL performance. Here, Factor denotes the inclusion of structured data, such as key clinical indicators extracted from electronic health records. Results show that introducing structured data alongside image and corpus inputs consistently improves performance across all metrics. Notably, the combination of all three modalities, image, corpus, and structured factors, achieves the highest scores, highlighting the complementary role of structured data in enhancing both accuracy and completeness of generated glaucoma diagnostic reports.

\begin{table}[h]
\centering
\caption{Impact of Input Modality ($\times$100\%)} 
\scalebox{0.87}{
\begin{tabular}{ccc|cccccc}
\hline
 \multicolumn{3}{c|}{\textbf{Data Category}}   & \multicolumn{6}{c}{\textbf{DA-SPL}}  \\ 
\hline
 \textbf{Image} & \multicolumn{1}{|c|}{\textbf{Corpus}} & \textbf{Factor} & \textbf{B-1} &  \textbf{B-2} & \textbf{B-3} & \textbf{B4}  & \textbf{ROU} & \textbf{CID} \\
\hline
\ding{55}  & \ding{55} & \ding{55} & - & - & - & - & - & -   \\
\ding{55}  & \ding{55} & \ding{51} & 5.12  & 4.96  & 5.08  & 5.21  & 4.87  & 5.03  \\
\ding{55}  & \ding{51} & \ding{55} & 15.20 & 10.35 & 7.14  & 6.23  & 14.26 & 20.35 \\
\ding{55}  & \ding{51} & \ding{51} & 25.45 & 16.82 & 11.97 & 8.92  & 22.73 & 31.84 \\
\ding{51}  & \ding{55} & \ding{55} & 35.68 & 23.21 & 16.43 & 12.25 & 31.42 & 43.72 \\
\ding{51}  & \ding{55} & \ding{51} & \underline{51.34} & \underline{33.42} & \underline{23.56} & \underline{17.42} & \underline{42.86} & \underline{58.03} \\
\ding{51}  & \ding{51} & \ding{55} & 45.12 & 29.54 & 20.89 & 15.63 & 38.92 & 52.18 \\
\ding{51}  & \ding{51} & \ding{51} & \textbf{56.97} & \textbf{36.71} & \textbf{25.14} & \textbf{18.74} & \textbf{45.08} & \textbf{62.44}   \\
\hline
\end{tabular}}
\label{ab-4}
\end{table}

\subsection{Supplementary Experiment}
Table~\ref{se} compares the original models with their improved counterparts after incorporating the proposed weight balance function (Eq.~\ref{eq7}). Across all architectures, the enhanced versions (\ding{51}) achieve consistent improvements over their baselines (\ding{55}) in BLEU, ROUGE-L, and CIDEr scores. The gains are most pronounced for transformer-based and attention-augmented models, such as AoANet and JE-TriNet, indicating that the weight balance function effectively optimizes feature contributions within these architectures. Although the performance of JE-TriNet (plus) remains slightly below that of DA-SPL, the observed improvements confirm the broad applicability and effectiveness of the proposed method across different model backbones.

\begin{table}[h]
\centering
\caption{Comparison of Models \& Improved Ones (×100\%)}
\label{se}
\scalebox{0.95}{
\begin{tabular}{l|c|cccccc}
\hline
\textbf{Model} & \textbf{Ver.}  & \textbf{B-1} &  \textbf{B-2} & \textbf{B-3} & \textbf{B4}  & \textbf{ROU} & \textbf{CID} \\
\hline
CNN-RNN     & \ding{55} & 42.13 & 26.34 & 17.25 & 11.28 & 36.27 & 32.41 \\
CNN-RNN     & \ding{51} & 43.58 & 27.46 & 17.92 & 11.85 & 37.42 & 33.76 \\
\hline
CNN-LSTM    & \ding{55} & 46.52 & 29.18 & 19.46 & 13.47 & 38.49 & 37.95 \\
CNN-LSTM    & \ding{51} & 47.96 & 30.27 & 20.12 & 14.03 & 39.64 & 39.24 \\
\hline
CaptionNet  & \ding{55} & 49.27 & 31.56 & 21.38 & 14.52 & 40.23 & 45.28 \\
CaptionNet  & \ding{51} & 50.63 & 32.64 & 22.04 & 15.08 & 41.38 & 46.73 \\
\hline
AoANet      & \ding{55} & 52.79 & 33.82 & 23.74 & 16.25 & 43.56 & 56.31 \\
AoANet      & \ding{51} & 54.21 & 34.91 & 24.38 & 16.81 & 44.71 & 57.82 \\
    \hline
    JE-TriNet   & \ding{55} & 54.38 & 36.17 & 25.49 & 17.93 & 44.87 & 61.84 \\
    JE-TriNet   & \ding{51} & 55.46 & 36.54 & 25.88 & 18.32 & 45.03 & 62.10 \\
\hline
\end{tabular}}
\end{table}

\subsection{Visualization}

Fig.~\ref{fig:r} presents two representative case studies from our report generation system. The first case (Panel a) demonstrates a high-risk glaucoma presentation, exhibiting characteristic findings including an enlarged optic disc (cup-to-disc ratio 0.8), diffuse neuroretinal rim thinning (most prominent in inferior and superior quadrants), ISNT rule violation, rim pallor, and associated structural changes. The second case (Panel b) illustrates a healthy control, displaying preserved neuroretinal rim architecture, absence of pathological disc features, and high confidence (95\%) in low-risk classification. Comparative analysis reveals that DA-SPL generates reports with improved conciseness and simplified sentence structures compared to ground truth annotations, while maintaining diagnostic accuracy. This optimization enhances clinical utility by improving readability and facilitating rapid interpretation.

\section{Conclusion}

We present DA-SPL, an innovative dual-attention semantic parallel-LSTM architecture for automated glaucoma diagnostic reporting from fundus imaging. DA-SPL's three key innovations: joint encoder attention mechanisms, parallelized LSTM decoder structure, and label enhancement module, collectively enable precise capture of subtle pathological features while optimizing semantic coherence and clinical relevance. The accompanying multimodal glaucoma corpus, comprising fundus images, structured clinical features, and expert-curated reports, represents a significant contribution to medical AI research infrastructure. Comprehensive evaluations across multiple benchmarks confirm DA-SPL's superior performance over existing methods in terms of both quantitative metrics and qualitative clinical utility, generating concise yet diagnostically precise reports that closely mirror expert annotations.

\section{Future Work}

Our ongoing research will focus on expanding and standardizing the proposed glaucoma-specific multimodal corpus, with planned public release to support broader scientific investigation. Future directions include: 
\begin{itemize}
    \item Systematic Integration of LLMs to Enhance Diagnostic Reasoning Capability
    \item Development of Advanced Multimodal Fusion Technique for Improved Feature Representation
    \item Clinical Validation Studies to Assess Real-world Applicability
\end{itemize}

We anticipate that combining comprehensive multimodal data resources with cutting-edge AI methodologies will yield transformative advances in glaucoma diagnosis and management, ultimately bridging the gap between computational research and clinical practice.

\section*{Acknowledgement}
This research work is partially sponsored by NASA under Grant No. 80NSSC22K0144, UTSW under GMO No. 241213, and the National Institutes of Health (NIH) under Grant No. 1R01AG083179-01.

\bibliography{main}{}

\begin{thebibliography}{10}

\bibitem{gpt4o}
Josh Achiam, Steven Adler, Sandhini Agarwal, Lama Ahmad, Ilge Akkaya, Florencia~Leoni Aleman, Diogo Almeida, Janko Altenschmidt, Sam Altman, Shyamal Anadkat, et~al.
\newblock Gpt-4 technical report.
\newblock {\em arXiv preprint arXiv:2303.08774}, 2023.

\bibitem{llama3}
Meta AI.
\newblock Llama 3 technical report.
\newblock \url{https://ai.meta.com/llama/}, 2024.
\newblock Accessed: August 2025.

\bibitem{bg-1}
{Dominic M.} Choo, Harper Henderson, Mahija Ginjupalli, and {Karanjit S.} Kooner.
\newblock {\em Glaucoma and retinal vein occlusion}, pages 11--38.
\newblock Nova Science Publishers, Inc., February 2025.

\bibitem{convit}
Stéphane d'Ascoli, Hugo Touvron, Matthew Leavitt, Ari Morcos, Giulio Biroli, and Levent Sagun.
\newblock Convit: Improving vision transformers with soft convolutional inductive biases.
\newblock In {\em International Conference on Machine Learning (ICML)}, pages 2286--2296. PMLR, 2021.

\bibitem{imagenet_cvpr09}
Jia Deng, Wei Dong, Richard Socher, Li-Jia Li, Kai Li, and Li~Fei-Fei.
\newblock Imagenet: A large-scale hierarchical image database.
\newblock In {\em Proceedings of the IEEE Conference on Computer Vision and Pattern Recognition (CVPR)}, pages 248--255, 2009.

\bibitem{SCES}
A.~Diaz‑Pinto et~al.
\newblock Cnns for automatic glaucoma assessment using fundus images: An extensive validation.
\newblock {\em Biomedical Engineering Online}, 2019.

\bibitem{cnn-lstm}
Jeff Donahue, Lisa Anne~Hendricks, Sergio Guadarrama, Marcus Rohrbach, Subhashini Venugopalan, Kate Saenko, and Trevor Darrell.
\newblock Long-term recurrent convolutional networks for visual recognition and description.
\newblock In {\em Proceedings of the IEEE Conference on Computer Vision and Pattern Recognition (CVPR)}, pages 2625--2634, 2015.

\bibitem{vit}
Alexey Dosovitskiy, Lucas Beyer, Alexander Kolesnikov, Dirk Weissenborn, Xiaohua Zhai, Thomas Unterthiner, Mostafa Dehghani, Matthias Minderer, Georg Heigold, Sylvain Gelly, Jakob Uszkoreit, and Neil Houlsby.
\newblock An image is worth 16x16 words: Transformers for image recognition at scale.
\newblock In {\em International Conference on Learning Representations (ICLR)}, 2021.

\bibitem{rimone_dl}
Francisco~José Fumero~Batista, Tinguaro Diaz‐Alemán, José Sigut, Silvia Alayón, Rafael Arnay, and Denisse Angel‑Pereira.
\newblock Rim‑one dl: A unified retinal image database for assessing glaucoma using deep learning.
\newblock {\em Image Analysis \& Stereology}, 39(3):161--167, 2020.

\bibitem{r1}
Daya Guo, Dejian Yang, Haowei Zhang, Junxiao Song, Ruoyu Zhang, Runxin Xu, Qihao Zhu, Shirong Ma, Peiyi Wang, Xiao Bi, et~al.
\newblock Deepseek-r1: Incentivizing reasoning capability in llms via reinforcement learning.
\newblock {\em arXiv preprint arXiv:2501.12948}, 2025.

\bibitem{resnet}
Kaiming He, Xiangyu Zhang, Shaoqing Ren, and Jian Sun.
\newblock Deep residual learning for image recognition.
\newblock In {\em Proceedings of the IEEE Conference on Computer Vision and Pattern Recognition (CVPR)}, pages 770--778. IEEE, 2016.

\bibitem{gelu}
Dan Hendrycks and Kevin Gimpel.
\newblock Gaussian error linear units (gelus).
\newblock {\em arXiv preprint arXiv:1606.08415}, 2016.

\bibitem{lstm}
Sepp Hochreiter and J{\"u}rgen Schmidhuber.
\newblock Long short-term memory.
\newblock {\em Neural Computation}, 9(8):1735--1780, 1997.

\bibitem{STARE}
Gary~A. Hoover and Michael Goldbaum.
\newblock Locating the optic nerve in a retinal image using the fuzzy convergence of the blood vessels.
\newblock {\em IEEE Transactions on Medical Imaging}, 22(8):951--958, 2003.

\bibitem{svaml}
Cheng Huang, Junhao Shen, Beichen Hu, Mohammad Ausaf Ali~Haqqani, Tsengdar Lee, Karanjit Kooner, Ning Zhang, and Jia Zhang.
\newblock Semantic and visual attention‑driven multi‑lstm network for automated clinical report generation.
\newblock In Arash Shaban‑Nejad, Martin Michalowski, and Simone Bianco, editors, {\em AI for Health Equity and Fairness: Leveraging AI to Address Social Determinants of Health}, volume 1164 of {\em Studies in Computational Intelligence}, pages 233--248. Springer, 2024.

\bibitem{RetinaMix}
Cheng Huang, Weizheng Xie, Fan Gao, Yutong Liu, Ruoling Wu, Zeyu Han, Jingxi Qiu, Xiangxiang Wang, Zhenglin Yang, Hao Wang, et~al.
\newblock Biovessel-net and retinamix: Unsupervised retinal vessel segmentation from octa images.
\newblock {\em arXiv preprint arXiv:2509.23617}, 2025.

\bibitem{glaboost}
Cheng Huang, Weizheng Xie, Karanjit Kooner, Tsengdar Lee, Jui-Kai Wang, and Jia Zhang.
\newblock Glaboost: A multimodal structured framework for glaucoma risk stratification.
\newblock {\em arXiv preprint arXiv:2508.03750}, 2025.

\bibitem{glagan}
Cheng Huang, Weizheng Xie, Tsengdar~J Lee, Jui-Kai Wang, Karanjit Kooner, Ning Zhang, and Jia Zhang.
\newblock X-gan: A generative ai-powered unsupervised model for main vessel segmentation of glaucoma screening, 2025.

\bibitem{glalstm}
Cheng Huang, Weizheng Xie, Jian Zhou, Tsengdar Lee, Karanjit Kooner, and Jia Zhang.
\newblock Glalstm: A concurrent lstm stream framework for glaucoma detection via biomarkers mining, 2025.
\newblock Accepted at IEEE EMBC 2025; arXiv preprint 2408.15555.

\bibitem{9478347}
Cheng Huang, Anyuan Yu, Yiwen Wang, and Honglin He.
\newblock Skin lesion segmentation based on mask r-cnn.
\newblock In {\em 2020 International Conference on Virtual Reality and Visualization (ICVRV)}, pages 63--67, 2020.

\bibitem{icct}
Cheng Huang, Yongbin Yu, and Minhui Qi.
\newblock Skin lesion segmentation based on deep learning.
\newblock In {\em 2020 IEEE 20th International Conference on Communication Technology (ICCT)}, pages 1360--1364, 2020.

\bibitem{aoa}
Lun Huang, Wenmin Wang, Jie Chen, and Xiao-Yong Wei.
\newblock Attention on attention for image captioning.
\newblock In {\em Proceedings of the IEEE/CVF International Conference on Computer Vision (ICCV)}, October 2019.

\bibitem{bg-2}
Yifan Huang, Dmitry Plotnikov, Han Wang, et~al.
\newblock Gwas-by-subtraction reveals an iop-independent component of primary open angle glaucoma.
\newblock {\em Nature Communications}, 15:8962, 2024.

\bibitem{cnn}
Alex Krizhevsky, Ilya Sutskever, and Geoffrey~E Hinton.
\newblock Imagenet classification with deep convolutional neural networks.
\newblock In {\em Advances in Neural Information Processing Systems (NeurIPS)}, volume~25, 2012.

\bibitem{lag}
Liu Li, Mai Xu, Hanruo Liu, Yang Li, Xiaofei Wang, Lai Jiang, Zulin Wang, Xiang Fan, and Ningli Wang.
\newblock A large-scale database and a cnn model for attention-based glaucoma detection.
\newblock {\em IEEE transactions on medical imaging}, 39(2):413--424, 2019.

\bibitem{lag2}
Liu Li, Mai Xu, Xiaofei Wang, Lai Jiang, and Hanruo Liu.
\newblock Attention based glaucoma detection: A large-scale database and cnn model.
\newblock In {\em Proceedings of the IEEE/CVF conference on computer vision and pattern recognition}, pages 10571--10580, 2019.

\bibitem{octa500}
Mingchao Li, Kun Huang, Qiuzhuo Xu, Jiadong Yang, Yuhan Zhang, Zexuan Ji, Keren Xie, Songtao Yuan, Qinghuai Liu, and Qiang Chen.
\newblock Octa-500: a retinal dataset for optical coherence tomography angiography study.
\newblock {\em Medical image analysis}, 93:103092, 2024.

\bibitem{rouge}
Chin-Yew Lin.
\newblock Rouge: A package for automatic evaluation of summaries.
\newblock In {\em Text Summarization Branches Out: Proceedings of the ACL-04 Workshop}, pages 74--81, 2004.

\bibitem{aensi}
Yihan Lin, Qian Tang, Hao Wang, Cheng Huang, Ekong Favour, Xiangxiang Wang, Xiao Feng, and Yongbin Yu.
\newblock Attention enhanced network with semantic inspector for medical image report generation.
\newblock In {\em 2023 IEEE 35th International Conference on Tools with Artificial Intelligence (ICTAI)}, pages 242--249, 2023.

\bibitem{v3}
Aixin Liu, Bei Feng, Bing Xue, Bingxuan Wang, Bochao Wu, Chengda Lu, Chenggang Zhao, Chengqi Deng, Chenyu Zhang, Chong Ruan, et~al.
\newblock Deepseek-v3 technical report.
\newblock {\em arXiv preprint arXiv:2412.19437}, 2024.

\bibitem{adamw}
Ilya Loshchilov and Frank Hutter.
\newblock Decoupled weight decay regularization.
\newblock {\em International Conference on Learning Representations (ICLR)}, 2019.
\newblock arXiv:1711.05101.

\bibitem{fairvision}
Yan Luo, Muhammad~Osama Khan, Yu~Tian, Min Shi, Zehao Dou, Tobias Elze, Yi~Fang, and Mengyu Wang.
\newblock Fairvision: Equitable deep learning for eye disease screening via fair identity scaling, 2024.

\bibitem{fairclip}
Yan Luo, Min Shi, Muhammad~Osama Khan, Muhammad~Muneeb Afzal, Hao Huang, Shuaihang Yuan, Yu~Tian, Luo Song, Ava Kouhana, Tobias Elze, Yi~Fang, and Mengyu Wang.
\newblock Fairclip: Harnessing fairness in vision-language learning.
\newblock In {\em Proceedings of the IEEE/CVF Conference on Computer Vision and Pattern Recognition (CVPR)}, 2024.

\bibitem{gdp}
Yan Luo, Min Shi, Yu~Tian, Tobias Elze, and Mengyu Wang.
\newblock Harvard glaucoma detection and progression: A multimodal multitask dataset and generalization‑reinforced semi‑supervised learning.
\newblock In {\em Proceedings of the IEEE/CVF International Conference on Computer Vision (ICCV)}, pages 20471--20482, 2023.

\bibitem{gf}
Yan Luo, Yu~Tian, Min Shi, Louis R. Pasquale, Lucy Q. Shen, Nazlee Zebardast, Tobias Elze, and Mengyu Wang.
\newblock Harvard glaucoma fairness: A retinal nerve disease dataset for fairness learning and fair identity normalization.
\newblock {\em IEEE Transactions on Medical Imaging}, 43(7):2623--2633, 2024.

\bibitem{rose}
Yuhui Ma, Huaying Hao, Huazhu Fu, Jiong Zhang, Jianlong Yang, Jianyang Yang, Jiang Liu, Yalin Zheng, and Yitian Zhao.
\newblock Rose: A retinal {OCT‐A}ngiography vessel segmentation dataset and new model.
\newblock {\em IEEE Transactions on Medical Imaging}, 40(3):928--939, 2021.

\bibitem{o3}
OpenAI.
\newblock Gpt-4o: Openai’s multimodal flagship model.
\newblock \url{https://openai.com/index/gpt-4o}, 2024.
\newblock Model ID: gpt-4o (o3); Accessed: August 1, 2025.

\bibitem{bleu}
Kishore Papineni, Salim Roukos, Todd Ward, and Wei-Jing Zhu.
\newblock Bleu: a method for automatic evaluation of machine translation.
\newblock In {\em Proceedings of the 40th Annual Meeting of the Association for Computational Linguistics}, pages 311--318, 2002.

\bibitem{mured}
Manuel~Alejandro Rodr{\'\i}guez, Hasan Al‑Marzouqi, and Panos Liatsis.
\newblock Multi‑label retinal disease classification using transformers.
\newblock {\em IEEE Journal of Biomedical and Health Informatics}, 2022.
\newblock Using the MuReD dataset (2,208 fundus images, 20 labels including glaucoma).

\bibitem{unet}
Olaf Ronneberger, Philipp Fischer, and Thomas Brox.
\newblock U-net: Convolutional networks for biomedical image segmentation.
\newblock In {\em Medical Image Computing and Computer-Assisted Intervention (MICCAI)}, volume 9351, pages 234--241. Springer, 2015.

\bibitem{drishti}
J.~Sivaswamy, G.~D. Joshi, and S.~R. Krishnadas.
\newblock Drishti‑gs: Retinal image dataset for optic nerve head (onh) segmentation.
\newblock In {\em IEEE 11th International Symposium on Biomedical Imaging (ISBI)}, pages 53--56. IEEE, 2014.

\bibitem{ridge}
Joes Staal, Michael~D. Abr{\`a}moff, Meindert Niemeijer, Max~A. Viergever, and Bram Van~Ginneken.
\newblock Ridge-based vessel segmentation in color images of the retina.
\newblock {\em IEEE Transactions on Medical Imaging}, 23(4):501--509, 2004.

\bibitem{fairseg}
Yu~Tian, Min Shi, Yan Luo, Ava Kouhana, Tobias Elze, and Mengyu Wang.
\newblock Fairseg: A large‑scale medical image segmentation dataset for fairness learning using segment anything model with fair error‑bound scaling.
\newblock In {\em International Conference on Learning Representations (ICLR)}, 2024.
\newblock Accepted as an ICLR 2024 conference paper.

\bibitem{fairdomain}
Yu~Tian, Congcong Wen, Min Shi, Muhammad~Muneeb Afzal, Hao Huang, Muhammad~Osama Khan, Yan Luo, Yi~Fang, and Mengyu Wang.
\newblock Fairdomain: Achieving fairness in cross-domain medical image segmentation and classification.
\newblock In {\em European Conference on Computer Vision (ECCV)}, 2024.

\bibitem{llama}
Hugo Touvron, Thibaut Lavril, Gautier Izacard, Xavier Martinet, Marie-Anne Lachaux, Timoth{\'e}e Lacroix, Baptiste Rozi{\`e}re, Naman Goyal, Eric Hambro, Faisal Azhar, et~al.
\newblock Llama: Open and efficient foundation language models.
\newblock {\em arXiv preprint arXiv:2302.13971}, 2023.

\bibitem{llama2}
Hugo Touvron, Louis Martin, Kevin Stone, Peter Albert, Amjad Almahairi, Yasmine Babaei, Nikolay Bashlykov, Soumya Batra, Prajjwal Bhargava, Shruti Bhosale, et~al.
\newblock Llama 2: Open foundation and fine-tuned chat models.
\newblock {\em arXiv preprint arXiv:2307.09288}, 2023.

\bibitem{att}
Ashish Vaswani, Noam Shazeer, Niki Parmar, Jakob Uszkoreit, Llion Jones, Aidan~N Gomez, {\L}ukasz Kaiser, and Illia Polosukhin.
\newblock Attention is all you need.
\newblock In {\em Advances in Neural Information Processing Systems (NeurIPS)}, volume~30. Curran Associates, Inc., 2017.

\bibitem{cider}
Ramakrishna Vedantam, C~Lawrence~Zitnick, and Devi Parikh.
\newblock Cider: Consensus-based image description evaluation.
\newblock In {\em Proceedings of the IEEE Conference on Computer Vision and Pattern Recognition (CVPR)}, pages 4566--4575, 2015.

\bibitem{cnn-rnn}
Oriol Vinyals, Alexander Toshev, Samy Bengio, and Dumitru Erhan.
\newblock Show and tell: A neural image caption generator.
\newblock In {\em Proceedings of the IEEE Conference on Computer Vision and Pattern Recognition (CVPR)}, pages 3156--3164, 2015.

\bibitem{IFE}
Longyu Yang, Hanli Wang, Pengjie Tang, and Qinyu Li.
\newblock Captionnet: A tailor-made recurrent neural network for generating image descriptions.
\newblock {\em IEEE Transactions on Multimedia}, 23:835--845, 2021.

\bibitem{joint}
Yan Yang, Jun Yu, Jian Zhang, Weidong Han, Hanliang Jiang, and Qingming Huang.
\newblock Joint embedding of deep visual and semantic features for medical image report generation.
\newblock {\em IEEE Transactions on Multimedia}, 25:167--178, 2021.

\bibitem{origa}
Zhuo Zhang, Feng~Shou Yin, Jiang Liu, Wing~Kee Wong, Ngan~Meng Tan, Beng~Hai Lee, Jun Cheng, and Tien~Yin Wong.
\newblock Origa(-light): An online retinal fundus image database for glaucoma analysis and research.
\newblock In {\em Annual International Conference of the IEEE Engineering in Medicine and Biology Society (EMBC)}, pages 3065--3068, 2010.

\end{thebibliography}
\bibliographystyle{plain}

\end{document}